\RequirePackage{fix-cm}

\documentclass[twocolumn]{svjour3}          
\smartqed                                   

\usepackage{graphicx}
\usepackage{mathptmx}
\usepackage{amsmath,amssymb}
\usepackage{mathtools}
\usepackage[numbers,comma,sort&compress]{natbib}

\begin{document}

\title{Analytical evaluation of relativistic molecular integrals. I. Auxiliary functions}

\author{
Ali Ba{\u g}c{\i}
\and
Philip E. Hoggan
}

\institute{
A. Ba{\u g}c{\i} \at
Department of Physics, Faculty of Arts and Sciences, Pamukkale University 20017 Denizli, Turkey
\\
\email{ali.bagci@yahoo.com.tr}
\and
P. E. Hoggan \at
Institute Pascal, UMR 6602 CNRS, University Blaise Pascal, 24 avenue des Landais BP 80026, 63177 Aubiere Cedex, France
}

\date{Received: date / Accepted: date}
\maketitle

\begin{abstract}
The auxiliary functions provide efficient computation of integrals arising at the self-consistent field (SCF) level for molecules using Slater-type bases. This applies both in relativistic and non-relativistic electronic structure theory.
The relativistic molecular auxiliary functions derived in our previous paper [Phys. Rev. E 91, 023303 (2015)] are discussed here in detail. Two solution methods are proposed in the present study. The ill-conditioned binomial series representation formulae first, are replaced by convergent series representation for incomplete beta functions then, they are improved by inserting an extra parameter used to extend the domain of convergence. Highly accurate results can be achieved for integrals by the procedures discussed in the present study which also places no restrictions on quantum numbers in all ranges of orbital parameters. The difficulty of obtaining analytical relations associated with using non-integer Slater-type orbitals which are non-analytic in the sense of complex analysis at $r=0$ is therefore, eliminated.
\end{abstract}

\section{Introduction}\label{intro}
When calculating molecular electronic structure at the Self-Consistent Field level(SCF), use of auxiliary functions in multi-center integral evaluation over Slater-type orbitals is one of the most efficient methods since it leads to fast and accurate calculations. It has a long history, beginning with Barnett and Coulson \cite{1_Coulson_1942,2_Barnett_1951}, Mulliken et al. \cite{3_Mulliken_1949}, Roothaan \cite{4_Roothaan_1951,5_Roothaan_1956}, R{\"u}denberg \cite{6_Rudenberg_1951}, L{\"o}wdin \cite{7_Lowdin_1956}, Kotani et al. \cite{8_Kotani_1963}, Harris and Mitchel \cite{9_Harris_1965,10_Harris_1966,11_Harris_1967}, Guseinov \cite{12_Guseinov_1970}. It is still being studied in the literature. In particular, the relationships obtained are constantly updated using developments in mathematical physics, chemistry and computer sciences \cite{13_Guseinov_2001,14_Harris_2002,15_Guseinov_2002,
16_Harris_2003,17_Harris_2004,18_Guseinov_2005,19_Fernandez_2006,
21_Guseinov_2009,20_Ema_2008,22_Lesiuk_2014,23_Lesiuk_2014,
24_Bagci_2014,25_Bagci_2015,26_Bagci_2015}.\\
Slater-type orbitals (STOs) \cite{27_Slater_1930,28_Parr_1957} are defined as follows:
\begin{equation} \label{eq:1}
\chi_{nlm} \left(\zeta, \textbf{\emph{r}}\right)=
\frac{\left(2\zeta \right)^{n+1/2}}{\sqrt{\Gamma(2n+1)}}r^{n-1}e^{-\zeta r}Y^{m}_{l}(\theta,\vartheta),
\end{equation}
here, $Y^{m}_{l}$ are complex or real spherical harmonics $(Y^{m*}_{l}=Y^{-m}_{l}; Y^{m}_{l} \equiv S^{m}_{l})$ differs from the Condon$-$Shortley phases by sign factor $(-1)^{m}$ \cite{29_Condon_1935, 30_Steinborn_1973}, $\left\lbrace n, l, m \right\rbrace$ are the principal, orbital, magnetic quantum numbers with, $n \in \mathbb{R}^{+}$, $0\leq l \leq \lfloor n \rfloor-1$, $-l \leq m \leq l$; $\lfloor n \rfloor$ stands for the integer part of $n$ and $\zeta$ are orbital parameters. They are obtained by simplification of Laguerre polynomial in hydrogen-like one-electron eigen-functions \cite{31_Willock_2009} by keeping only the term of the highest power of $r$. Other exponentially decaying bases, including the hydrogen-like functions may be written as linear combinations of STOs \cite{32_Weniger_2002}.
The related eigenfunctions satisfy Kato's cusp conditions for asymptotic behavior of the wave function near the nucleus \cite{33_Kato_1957} at long range all these orbitals  decrease exponentially \cite{34_Agmon_1982}. They are, therefore the natural choice of basis orbital in algebraic solution of the Schr{\"o}dinger equation for many-electron systems. They play a key role in the understanding of quantum mechanical theoretical problems; arising from testing limits of the methods obtained approximately to represent the physical systems, where reliable description of electron density is important e.g., investigation the molecules under strong magnetic field \cite{35_Bouferguene_1999}, second order perturbation energy corrections \cite{36_Hoggan_2011}.\\
So far, auxiliary functions for evaluation of molecular integrals over the STOs have been derived only for the integer values of principal quantum numbers. In that case, $n \in \mathbb{Z}^{+}$, $0\leq l \leq n-1$ and $\Gamma\left( 2n +1 \right)=n!$. The use of non-integer principal quantum numbers in STOs however, promise better results because they provide extra flexibility for closer variational description of molecules \cite{28_Parr_1957}. The vital importance of generalising  auxiliary function methods is then clear when extending the domain of applications that were previously limited to atoms \cite{37_Koga_1997,38_Koga_1997,39_Koga_2000,40_Guseinov_2012}, is considered. A basis spinor to be used in relativistic electronic structure calculation is obtained from the hydrogen atom Dirac-Hamiltonian eigenfunctions and it can be written in terms of Slater-type orbitals since its radial part has the following form \cite{41_Grant_2007,42_Bagci_2016}:
\begin{equation}\label{eq:2}
f(\zeta,r)
=\left\{{Ar^{n}+\zeta Br^{n+1}}\right\}e^{-\zeta r}.
\end{equation}
Deriving such mathematical tools therefore directly helps to evaluate the integrals arising from algebraic solution of the molecular Dirac equation since they in turn reduce to integrals over STOs.

The authors in their recent study \cite{25_Bagci_2015} through Laplace expansion of Coulomb interaction and prolate spheroidal coordinates, expressed the two-center molecular integrals in terms of so called relativistic molecular auxiliary functions. These auxiliary functions were evaluated numerically via Global-adaptive method with Gauss-Kronrod numerical integration extension. Note that highly accurate values from the suggested numerical method are available only in Mathematica programming language. Extended-precision Fortran is being investigated for applications, since the Mathematica programming language is suitable for bench-marking but requires prohibitive calculation time. One of us obtains analytical relations investigated in the reference, \cite{43_Bagci_2017}. The relativistic auxiliary functions are expressed through series representation of incomplete beta functions and in terms of integrals involving Appell functions. \\
Double series of Appell's functions \cite{44_Appell_1925},
\begin{equation}\label{eq:3}
F_{1} \left(a;b_{1},b_{2};c;x,y \right)
=\sum_{s_{1},s_{2}=0}^{\infty}
\frac{\left(a\right)_{s_{1}+s_{2}}\left(b_{1}\right)_{s_{1}}\left(b_{2}\right)_{s_{2}}}{\left(c\right)_{s_{1}s_{2}}s_{1}!s_{2}!}x^{s_{1}}y^{s_{2}},
\end{equation}
with $\left(\alpha \right)_{n}$ is Pochammer symbol, is mathematically convergent when the variables $\left(x,y\right)$, $\vert x \vert <1$ and $\vert y \vert <1$. Since the variables $\left(x,y\right)$ arising in relativistic auxiliary functions have values outside of convergence region it is necessary to make use of recurrence relations formulae \cite{45_Wang_2012} or a numerical integration of a third order ordinary differential equation that represents the system of partial differential equations of Appell functions given for a set of analytic continuations \cite{46_Colavecchia_2001}. These methods are, however, computationally inefficient and may not give correct results for a particular set of parameters $\left\lbrace a, b_{1}, b_{2},c \right\rbrace$. Besides, computing the Appell's functions without erroneous last digits is still being studied in the literature \cite{47_Colavecchia_2004}.

In the present study, we refer to the introductory remarks given in previous studies. Certain concepts and the results of previous work are also used. Here, the relationships given in terms of integrals involving Appell's functions are also reduced to series representation formulae for incomplete beta functions. Computing Appell's functions is therefore avoided. Furthermore, a new binomial expansion method is developed through that given in \cite{48_Liao_2004,49_Liu_2010}, where an extra parameter is used to extend the domain of convergence of the well-known Newton binomial expansion approximation. The ill-conditioned binomial series representation used for evaluation of the molecular integrals in the literature \cite{21_Guseinov_2009,50_Mekelleche_2000,51_Guseinov_2002,52_Guseinov_2002} is thus improved. Therefore, reliable methods to analytically evaluate the molecular integrals over non-integer Slater-type orbitals are finally obtained in the present work.
\section{Evaluation of relativistic molecular auxiliary functions}\label{sec:1_eval_mol_aux}
The compact expressions we previously derived for two-center, one- and two-electron relativistic molecular integrals in a lined-up coordinate system through Laplace expansion of Coulomb interaction and prolate spheroidal coordinates ($\xi, \nu, \phi$) where, $1\leq\xi<\infty$, $-1\leq\nu\leq1$, $0\leq\phi\leq2\pi$, are obtained using the following auxiliary functions integrals \cite{25_Bagci_2015}:
\begin{multline} \label{eq:4}
\left\lbrace \begin{array}{cc}
\mathcal{P}^{n_1,q}_{n_{2}n_{3}n_{4}}\left(p_{123} \right)
\\
\mathcal{Q}^{n_1,q}_{n_{2}n_{3}n_{4}}\left(p_{123} \right)
\end{array} \right\rbrace
\\
=\frac{p_{1}^{\sl n_{1}}}{\left({\sl n_{4}}-{\sl n_{1}} \right)_{\sl n_{1}}}
\int_{1}^{\infty}\int_{-1}^{1}{\left(\xi\nu \right)^{q}\left(\xi+\nu \right)^{\sl n_{2}}\left(\xi-\nu \right)^{\sl n_{3}}}\\ \times
\left\lbrace \begin{array}{cc}
P\left[{\sl n_{4}-n_{1}},p_{1}(\xi+\nu) \right]
\\
Q\left[{\sl n_{4}-n_{1}},p_{1}(\xi+\nu) \right]
\end{array} \right\rbrace
e^{p_{2}\xi-p_{3}\nu}d\xi d\nu,
\end{multline}
here, $\left\lbrace q, n_{1} \right\rbrace \in \mathbb{Z}$, $\left\lbrace n_{2}, n_{3}, n_{4}\right\rbrace \in \mathbb{R}$, $p_{123}=\left\lbrace p_{1}, p_{2}, p_{3}\right\rbrace$ (and in subsequent notations). $P,Q$ are the normalized complementary incomplete gamma and the normalized incomplete gamma functions,
\begin{equation}\label{eq:5}
P\left[\alpha, z \right]
=\frac{\gamma\left(\alpha,z\right)}{\Gamma\left(\alpha\right)},
\hspace{5mm}
Q\left[\alpha, z \right]
=\frac{\Gamma\left(\alpha,z\right)}{\Gamma\left(\alpha\right)},
\end{equation}
with $\gamma(a,z)$ and $\Gamma(a,z)$ are incomplete gamma functions,
\begin{equation}\label{eq:6}
\gamma\left(\alpha, z\right)
=\int_{0}^{z} t^{\alpha-1}e^{-t}dt,
\hspace{5mm}
\Gamma\left(\alpha, z\right)
=\int_{z}^{\infty} t^{\alpha-1}e^{-t}dt,
\end{equation}
$\Gamma(a)$ is a complete gamma function,
\begin{equation}\label{eq:7}
\Gamma\left(\alpha\right)
=\Gamma\left(\alpha, z\right)+\gamma\left(\alpha, z\right),
\end{equation}
and the Pochhammer symbol $(\alpha)_{n}$,
\begin{equation}\label{eq:8}
\left(\alpha\right)_{n}
=\frac{\Gamma\left(\alpha+n\right)}{\Gamma\left(\alpha\right)},
\end{equation}
respectively \cite{53_Abramowitz_1972, 54_Temme_1994}. Evaluation of these auxiliary functions involve some challenges including power functions with non-integer exponents. Also, incomplete gamma functions and their products have no explicit closed-form relations. On the other hand, symmetry properties of two-center two-electron integrals allow us to take advantage of sum $P+Q=1$ and so, represent the Eq. (\ref{eq:4}) through upward and downward distant recurrence relations of normalized incomplete gamma functions
\begin{multline}\label{eq:9}
\left\lbrace \begin{array}{cc}
P\left[a,bz \right]
\\
Q\left[a,bz \right]
\end{array} \right\rbrace
\\
=
\left\lbrace \begin{array}{cc}
P\left[a+n,bz \right]+e^{-bz}\sum_{s=1}^{n}\frac{\left(bz\right)^{a+s-1}}{\Gamma\left(a+s \right)}
\\
Q\left[a+n,bz \right]-e^{-bz}\sum_{s=1}^{n}\frac{\left(bz\right)^{a+s-1}}{\Gamma\left(a+s \right)}
\end{array} \right\rbrace,
\end{multline}
\begin{multline}\label{eq:10}
\left\lbrace \begin{array}{cc}
P\left[a,bz \right]
\\
Q\left[a,bz \right]
\end{array} \right\rbrace
\\
=
\left\lbrace \begin{array}{cc}
P\left[a-n,bz \right]-e^{-bz}\sum_{s=1}^{n-1}\frac{\left(bz\right)^{a-s-1}}{\Gamma\left(a-s \right)}
\\
Q\left[a-n,bz \right]+e^{-bz}\sum_{s=1}^{n-1}\frac{\left(bz\right)^{a-s-1}}{\Gamma\left(a-s \right)}
\end{array} \right\rbrace,
\end{multline}
in terms of the following form (please see \cite{43_Bagci_2017}):
\begin{multline}\label{eq:11}
\mathcal{G}^{n_{1},q}_{n_{2}n_{3}}\left(p_{123} \right)
=\frac{p_{1}^{n_{1}}}{\Gamma\left(n_{1}+1 \right)}
\\
\times \int_{1}^{\infty}\int_{-1}^{1} \left(\xi\nu\right)^{q}\left(\xi+\nu\right)^{n_{2}}\left(\xi-\nu \right)^{n_{3}}e^{-p_{2}\xi-p_{3}\nu}d\xi d\nu.
\end{multline}
The feature given above can therefore generally be defined as follows:

\textbf{Criterion.} Let $P\left[n_{4}-n_{1}, z \right]$ and $Q \left[n'_{4}-n'_{1}, z \right]$ then $n_{4}-n_{1}=a \pm c$, $n'_{4}-n'_{1}=a \pm d$, where $a \in \mathbb{R}$, $\left\lbrace c, d \right\rbrace \in \mathbb{Z}$ are true for any integrals that can be reduced to Eq. (\ref{eq:4}).
\vspace{2mm}\\
\textbf{Case 1.} The parameter $p_{3}=0$.\\
Starting by lowering the indices $q$ using,
\begin{equation}\label{eq:12}
\left(\xi\nu \right)=\frac{1}{4}\left\lbrace \left(\xi+\nu \right)^{2}-\left(\xi-\nu \right)^{2} \right\rbrace,
\end{equation}
the auxiliary functions $\mathcal{G}^{n_{1},q}$ are obtained as follows \cite{43_Bagci_2017},
\begin{equation}\label{eq:13}
\mathcal{G}^{n_{1},q}_{n_{2}n_{3}}\left(p_{120} \right)
=\frac{1}{4}\left\lbrace \mathcal{G}^{n_{1},q-1}_{n_{2}+2n_{3}}\left(p_{123} \right)-\mathcal{G}^{n_{1},q-1}_{n_{2}n_{3}+2}\left(p_{123} \right) \right\rbrace,
\end{equation}
here,
\begin{multline}\label{eq:14}
\mathcal{G}^{n_{1},0}_{n_{2}n_{3}}\left(p_{120} \right)
=h^{n_{1},0}_{n_{2}n_{3}}\left(p_{12} \right)+ h^{n_{1},0}_{n_{3}n_{2}}\left(p_{12}\right)
\\
-k^{n_{1},0}_{n_{2}n_{3}}\left(p_{12}\right)
-k^{n_{1},0}_{n_{3}n_{2}}\left(p_{12}\right),
\end{multline}
\begin{multline}\label{eq:15}
h^{n_{1},q'}_{n_{2}n_{3}}\left(p_{12}\right)
=\frac{p_{1}^{n_{1}}}{\Gamma\left(n_{1}+1 \right)}2^{n_{2}+n_{3}+q'+1}B\left(n_{2}+1,n_{3}+1\right)
\\
\times E_{-\left(n_{2}+n_{3}+q'+1\right)}\left(p_{2}\right)
-l^{n_{1},q'}_{n_{2}n_{3}}\left(p_{12}\right),
\end{multline}
\begin{multline}\label{eq:16}
l^{n_{1},q'}_{n_{2}n_{3}}\left(p_{12}\right)
\\
=\frac{p_{1}^{n_{1}}}{\Gamma\left(n_{1}+1\right)}
\sum_{s=0}^{\infty}\frac{\left(-n_{2}\right)_{s}}{\left(n_{3}+s+1\right)!}m^{n_{2}+q'-s}_{n_{3}+s+1}\left(p_{2}\right),
\end{multline}
\begin{equation}\label{eq:17}
m^{n_{1}}_{n_{2}}\left(p\right)
=2^{n_{1}}U\left(n_{2}+1,n_{1}+n_{2}+2,p\right)\Gamma\left(n_{2}+1\right)e^{-p},
\end{equation}
and,
\begin{multline}\label{eq:18}
k^{n_{1},q'}_{n_{2},n_{3}}\left(p_{12}\right)
=\frac{p_{1}^{n_{1}}}{\Gamma\left(n_{1}+1 \right)}2^{n_{2}+n_{3}+q'+1}B\left(n_{2}+1,n_{3}+1, \frac{1}{2}\right)
\\
\times E_{-\left(n_{2}+n_{3}+q'+1\right)}\left(p_{2}\right),
\end{multline}
with,
\begin{multline}\label{eq:19}
U\left(a,b;z\right)
=\frac{\Gamma\left(b-1\right)}{\Gamma\left(a\right)}{_{1}F_{1}}\left(a-b+1,2-b;z\right)
\\
+
\frac{\Gamma\left(1-b\right)}{\Gamma\left(a-b+1\right)}{_{1}F_{1}}\left(a;b;z\right),
\end{multline}
here, $U\left(a,b;z\right)$ are confluent hypergeometric functions of second kind with,
\begin{multline}\label{eq:20}
{_{1}F_{1}}\left(a,b;z\right)
\\
=\frac{\Gamma\left(b\right)}{\Gamma\left(b-a\right)\Gamma\left(a\right)}
\int_{0}^{1}t^{a-1}\left(1-t\right)^{b-a-1}e^{z t}dt,
\end{multline}
confluent hypergeometric functions of first kind and $B\left(a,b, z\right)$ incomplete beta functions,
\begin{equation}\label{eq:21}
B\left(a,b, z\right)
=\int_{0}^{z}t^{a-1}(1-t)^{b-1}dt,
\end{equation}
$B\left(a,b \right)=B\left(a,b,1\right)$ are beta functions, respectively \cite{53_Abramowitz_1972}.
\vspace{2mm}\\
\textbf{Case 2.} The parameter $p_{3} \ne 0$.\\
By lowering the indices $q$ and using the series expansion of exponential functions $e^{z}$, where $z=-p_{3} \nu$, the following relation is obtained \cite{43_Bagci_2017},
\begin{multline}\label{eq:22}
{\mathcal{G}^{n_{1},0}_{n_{2}n_{3}}}(p_{123})
=\frac{p_{1}^{n_{1}}}{\Gamma\left(n_{1}+1 \right)}
\sum_{s=0}^{\infty}
\frac{p_{3}^{s}}{\Gamma\left(s+1 \right)}
\left(\frac{1}{s+1}\right)
\\
\times
\left\lbrace
J_{n_{2}n_{3}}^{s+1,s+2;0}\left(p_{2}\right)
+\left(-1\right)^{s}J_{n_{3}n_{2}}^{s+1,s+2;0}\left(p_{2}\right)
\right\rbrace,
\end{multline}
The $J^{s,q}$ functions involve Appell's hypergeometric functions \cite{44_Appell_1925},
\begin{multline}\label{eq:23}
F_{1}\left(a;b_{1},b_{2};c;x,y \right)=
\frac{\Gamma\left(c\right)}{\Gamma\left(a\right)\Gamma\left(a-c\right)}\\
\times \int_{0}^{1}u^{a-1}\left(1-u\right)^{c-a-1}
\left(1-ux\right)^{-b_{1}}
\left(1-uy\right)^{-b_{2}}du,
\end{multline}
and their explicit forms are given as,
\begin{multline}\label{eq:24}
J_{n_{2}n_{3}}^{s,s';q}\left(p\right)
=\int_{1}^{\infty}F_{1}\left(s;-n_{2},-n_{3};s';\frac{1}{\xi},-\frac{1}{\xi} \right)\\
\times \xi^{n_{2}+n_{3}+q}e^{-p\xi}d\xi.
\end{multline}
The sum of two $J_{n_{2}n_{3}}^{s,s';q}$ functions arising in the right-hand-side of Eq.(\ref{eq:22}) is an integral in the form written as:
\begin{multline}\label{eq:25}
\left(\frac{1}{s+1}\right)
\left\lbrace
J_{n_{2}n_{3}}^{s+1,s+2;0}\left(p_{2}\right)
+\left(-1\right)^{s}J_{n_{3}n_{2}}^{s+1,s+2;0}\left(p_{2}\right)
\right\rbrace
\\
=\int_{1}^{\infty}\int_{-1}^{1}
\left(\xi+\nu\right)^{n_{2}}
\left(\xi-\nu\right)^{n_{3}}
\nu^{s}e^{-p_{2}\xi} d\xi d\nu.
\end{multline}
By dividing and multiplying the expression with $\xi^{s'}$ we obtain:
\begin{multline}\label{eq:26}
\mathcal{J}_{n_{2}n_{3}}^{s,s'}\left(p_{2}\right)
\\
=\int_{1}^{\infty}\int_{-1}^{1}
\left(\xi+\nu\right)^{n_{2}}
\left(\xi-\nu\right)^{n_{3}}
\left(\xi^{s'}\nu^{s} \right) \xi^{-s'} e^{-p_{2}\xi} d\xi d\nu.
\end{multline}
By again making use of Eq.(\ref{eq:12}), finally the following relation is obtained for $\left( s'=s \right)$:
\begin{equation}\label{eq:27}
\mathcal{J}^{s,s}_{n_{2}n_{3}}\left(p_{2} \right)
=\frac{1}{4}\left\lbrace \mathcal{J}^{s-1,s}_{n_{2}+2n_{3}}\left(p_{2} \right)
-\mathcal{J}^{s-1,s}_{n_{2}n_{3}+2}\left(p_{2} \right) \right\rbrace,
\end{equation}
\begin{multline}\label{eq:28}
\mathcal{J}_{n_{2}n_{3}}^{0,s}\left(p_{2}\right)
=\frac{1}{2^s}
\left\lbrace
h^{1,s}_{n_{2}n_{3}}\left(p_{02} \right)+ h^{1,s}_{n_{3}n_{2}}\left(p_{02}\right)
\right.
\\
\left.
-k^{1,s}_{n_{2}n_{3}}\left(p_{02}\right)
-k^{1,s}_{n_{3}n_{2}}\left(p_{02}\right)
\right\rbrace
\end{multline}
with, $p_{02}=\left\lbrace 1,p_{2} \right\rbrace$. It should be note that Eqs. (\ref{eq:14}, \ref{eq:28}) imply convergence properties of incomplete beta function expansions; $B_{z}\left(n_{1}, n_{2} \right)$ at $z=0$, where absolute value of $z$ must be $\vert z \vert < 1$. Considering the domain given for auxiliary functions $\mathcal{G}^{n_1,q}$, it is easy to see that the convergence condition is satisfied, here, $z=\frac{\xi-1}{2\xi}$. The Eq. (\ref{eq:22}) gives the convergence properties for series representation of exponential functions $e^{z}$ which are uniformly convergent for the entire complex plane for any $z$ with $\vert z \vert < \infty$.
\subsection{\textbf{On the use of Newton's binomial theorem}}\label{sec:binomG}
 Newton's binomial theorem is generalized by Liao, within the frame of the homotopic analysis \cite{48_Liao_2004}. An extra parameter $h$, the so-called auxiliary parameter is used to extend the domain of convergence. The auxiliary parameter is generally used in homotopic analysis to construct the so-called zero-order deformation equation. A set of expressions is thus obtained in terms of the auxiliary parameter $h$ as solutions.

 Series with the mean convergence domain show rate of solution increased by choosing a proper value for $h$ \cite{48_Liao_2004,49_Liu_2010}.

A power function such as $\left(\xi \pm \nu\right)^{n}$ with real number $n$ $(n \neq 0,1,2,3,...)$, can be written in a form that,
\begin{equation}\label{eq:29}
\left(\xi \pm \nu\right)^{n}
=\xi^{n}\left(1 \pm \frac{\nu_{0}}{\xi_{0}} \right)^{n}
\left(1 \pm \frac{\frac{\nu}{\xi}-\frac{\nu_{0}}{\xi_{0}}}{1 \pm \frac{\nu_{0}}{\xi_{0}}} \right)^{n},
\end{equation}
where, $\bigg\vert \frac{\frac{\nu}{\xi}-\frac{\nu_{0}}{\xi_{0}}}{1 \pm \frac{\nu_{0}}{\xi_{0}}} \bigg\vert\ <1$, $\frac{\nu_{0}}{\xi_{0}}= \mp 1 \mp \frac{1}{h}$ with $\vert \frac{\nu_{0}}{\xi_{0}} \vert<1$, respectively. The auxiliary parameter $h$ is then adjusted accordingly. By applying now the usual Newton's binomial expansion the following relations are obtained,
\begin{equation}\label{eq:30}
\left(\xi \pm \nu\right)^{n}
=\lim_{N\rightarrow \infty}
\sum_{s=0}^{N}
\left(\pm 1 \right)^s \mu_{n}^{N,s}\left(h\right)F_{s}\left(n\right)\xi^{n-s}\nu^{s},
\end{equation}
\begin{equation}\label{eq:31}
\mu_{n}^{N,s}\left(h\right)
=\sum_{s'=0}^{N-s}\left(\pm 1\right)^{s'}F_{s'}\left(n-s\right)
\left(-h\right)^{s-n}\left(h+1\right)^{s'}.
\end{equation}
The terms arising in Eq. (\ref{eq:4}) can thus be re-written as:
\begin{multline}\label{eq:32}
\left(\xi + \nu \right)^{n_{2}}
\left(\xi - \nu \right)^{n_{2}}
=\lim_{N \to \infty}
\sum_{s,s'=0}^{N}
\mu_{n_{2}}^{N,s}\left(h\right)
\mu_{n_{3}}^{N,s'}\left(h'\right)
\\
\times
F_{s}\left(n_{2}\right)
F_{s'}\left(n_{3}\right)
\xi^{n_{2}+n_{3}-s-s'}\nu^{s+s'},
\end{multline}
where, $F_{s}\left(n\right)$, are the binomial coefficients indexed by $n$, $s$ is usually written $\left(\begin{array}{cc}n\\s\end{array} \right)$, with,
\begin{equation}\label{eq:33}
\left(\begin{array}{cc}n\\s\end{array} \right)
=\frac{\Gamma\left(n+1\right)}{\Gamma\left(s+1\right)\Gamma\left(n-s+1\right)}.
\end{equation}
According to formulae given above the auxiliary functions $\mathcal{G}^{n_{1},q}$ are obtained as follows:
\begin{multline}\label{eq:34}
{\mathcal{G}^{n_{1},q}_{n_{2}n_{3}}}(p_{123})
=\frac{p_{1}^{n_{1}}}{\Gamma\left(n_{1}+1\right)}
\lim_{N \to \infty}\sum_{s,s'=0}^{N}
\mu_{n_{2}}^{N,s}\left(h\right)
\mu_{n_{3}}^{N,s'}\left(h'\right)
\\
\times
F_{s}\left(n_{2}\right)
F_{s'}\left(n_{3}\right)
\int_{1}^{\infty}\xi^{n_{1}+n_{2}+q-s-s'}e^{-p_{2} \xi}d\xi
\\
\times
\int_{-1}^{1}\nu^{q+s+s'}e^{-p_{3} \nu}d\nu,
\end{multline}
\begin{multline}\label{eq:35}
\mathcal{G}^{n_{1},q}_{n_{2}n_{3}}\left(p_{123} \right)
=\frac{p_{1}^{n_{1}}}
{\Gamma\left(n_{1}+1 \right)}
\lim_{N \to \infty}\sum_{s,s'=0}^{N}
\mu_{n_{2}}^{N,s}\left(h\right)
\mu_{n_{3}}^{N,s'}\left(h'\right)
\\
\times
F_{s}\left(n_{2}\right)
F_{s'}\left(n_{3}\right)
\left\lbrace \frac{E_{-\left(n_{2}+n_{3}\right)-q+k}\left(p_{2}\right)}{p_{3}^{q+k+1}} \right.\\
\left. \times \Bigg(\gamma\left(q+k+1,p_{3}\right)-\gamma\left(q+k+1,-p_{3}\right) \Bigg)\right\rbrace,
\end{multline}
where, $k=s+s'$ and,
\begin{equation}\label{eq:36}
E_{n}\left(p\right)
=\int_{1}^{\infty}\frac{e^{-p\xi}}{\xi^{n}}d\xi,
\end{equation}
are the exponential integral functions.
\section{Conclusion}\label{sec:conclusion}
The renewed interest in molecular integrals over Slater-type orbitals with non-integer principal quantum numbers is increasing. Recent studies show that they are used in both relativistic and non-relativistic electronic structure calculations. These integrals are expressed in terms of molecular auxiliary functions. They involve power functions such as $f\left(z \right)=z^n=e^{n\log z}$ with non-integer exponents $n \in \mathbb{R}$ which can not be represented by a power series because they are not analytic about $z=0$ \cite{55_Weniger_2008}. This constitutes the underlying reason why the Slater-type orbitals with non-integer principal quantum numbers could not be used in molecular electronic structure calculations so far. Availability of computation methods for molecular auxiliary functions on the other hand, need urgen implementation and are precious. Two methods based on this reasoning are proposed in this study. Firstly, through expansion of exponential functions the molecular $\mathcal{G}^{n,q}$ auxiliary functions reduce to integrals involving Appell functions (Eq.(\ref{eq:22})). Instead of using recurrence relations of Appell's functions, they are represented through convergent series expansion of incomplete beta functions. Secondly, through an improved form of the binomial series expansion of power functions they reduce to easily integrable expressions in which the variables are separated (Eq.(\ref{eq:35})). These methods are derived according to a criterion given below the Eq.(\ref{eq:11}). This criterion is correct, from using the Laplace expansion of the Coulomb interaction to evaluate multicenter integrals (please see \cite{43_Bagci_2017} for more detail). The relationships given in the presented work are reliable and convergent. Benchmark results in our previous papers \cite{24_Bagci_2014,25_Bagci_2015} can therefore be obtained with the formalism given in the present study.

The homotopy analysis method which is used to extend the domain of convergence of Newton's binomial series representation formulae may also be used to obtain non-analytic
solutions, which by their nature can not be expressed through the power series \cite{56_Gorder_2017}. The single-center expansion method i.e., expansion of Slater-type orbitals with non-integer principal quantum numbers in terms of an infinite series of (integer principal quantum number) Slater-type orbitals. \cite{57_Guseinov_2002,58_Guseinov_2007},
\begin{equation} \label{eq:37}
\chi_{n lm}(\zeta,\textbf{\emph{r}})
=\sum_{\mu=l+1}^{\infty} V_{n l,\mu' l}\chi_{\mu' lm}(\zeta,\textbf{\emph{r}}),
\end{equation}
where, $V$ are the expansion coefficients and $ \mu \in \mathbb{Z}^{+}$,  may thus also become available.

The computational aspect of the formulae given here for molecular auxiliary functions and their applications will be the subject of future research.
\section*{Acknowledgement}
A.B. would like to thank to Department of Physics, Faculty of Arts and Sciences, Pamukkale University for providing working facilities. He would like also give a special thank to Prof. Dr. Muzaffer Adak and Prof. Dr. Mestan Kalay for their fruitful discussions.

\end{document}